\documentclass[11pt]{article} 
\usepackage{a4,fullpage}
\usepackage{times}
\usepackage{url}
\usepackage{epsfig}
\usepackage{amssymb}

\newtheorem{mfproposition}{Proposition}
\usepackage{bbm}
\usepackage{amstext}
\usepackage{color}
\usepackage{xspace}
\usepackage{amsfonts}
\usepackage{amssymb}
\usepackage{graphicx}
\usepackage{accents}
\usepackage{scalefnt}
\DeclareRobustCommand\bfsf{\normalfont\bfseries\sf\boldmath}
\newcommand*{\bffotlx}{\hbox{\bfsf FOTLX}}
\newcommand*{\fotlx}{\textsf{FOTLX}\xspace}
\newcommand{\gb}{\mathfrak{b}}

\newcommand*{\imp}{\Rightarrow}

\newcommand*{\plus}{\ensuremath{+}}
\newcommand*{\minus}{\ensuremath{-}}

\newcommand{\const}{\ensuremath{\mathrm{const}}}
\newcommand{\next}{\!\raisebox{-.2ex}{ 
                        \mbox{\unitlength=0.9ex
                        \begin{picture}(2,2)
                        \linethickness{0.06ex}
                        \put(1,1){\circle{2}} 
                        \end{picture}}}       
                        \,}

         {{\hfill\tiny$\Box$}\par\vspace*{1ex}}
\newenvironment{proof}{\par\vspace*{0.5ex}\noindent\rm {\bf Proof:} }%
         {{\hfill\tiny$\Box$}\par\vspace*{1ex}}

\newtheorem{lemma}{Lemma}
\newtheorem{theorem}[lemma]{Theorem}

\newtheorem{definition}{Definition}
\newtheorem{example}{Example}

\newcommand{\true}{\textsc{true}}
\newcommand{\false}{\textsc{false}}

\newcommand{\uP}{\ensuremath{\mathcal{U}}}

\newcommand{\eP}{\ensuremath{\mathcal{E}}}
\newcommand{\sP}{\ensuremath{\mathcal{S}}}
\newcommand{\iP}{\ensuremath{\mathcal{I}}}

\newcommand{\mT}{\ensuremath{\mathcal{T}}}

\newcommand{\gM}{\mathfrak{M}}
\newcommand{\ga}{\mathfrak{a}}

\newcommand{\gA}{\mathfrak{V}}

\newcommand{\Nat}{\mathbbm{N}}
\newcommand{\sometime}{\lozenge}
\newcommand{\sometimes}{\lozenge}

\newcommand{\until}{\mathcal{U}}

\newcommand{\union}{\cup}

\def\true{\hbox{\bf true}}
\def\false{\hbox{\bf false}}
\def\start{\hbox{\bf start}}

\newcommand{\TProb}{\textsf{P}}

\newcommand{\implies}{\Rightarrow}

\newcommand*{\fotl}{\ensuremath{\mathsf{FOTL}}\xspace}

\newcommand*{\New}{\color{black}}

\makeatletter
        {\begin{list}{}{
              \setlength{\itemsep}{0pt}
                \settowidth{\labelwidth}{#1}
                \settowidth{\leftmargin}{#1\hspace{\labelsep}}}
        }{
        \end{list} }
\makeatother

\newcommand{\TRPppV}{\textbf{TRP\protect\raisebox{0.35ex}{\ensuremath{\scriptstyle{+}{+}\;}}V}\xspace}

\def\until{\hbox{$\,\mathsf{U} \,$}}
\def\unless{\hbox{$\,\mathsf{W} \,$}}

\newcommand{\always}{\raisebox{-.2ex}{
			   \mbox{\unitlength=0.9ex
			   \begin{picture}(2,2)
			   \linethickness{0.06ex}
			   \put(0,0){\line(1,0){2}}
			   \put(0,2){\line(1,0){2}}
			   \put(0,0){\line(0,1){2}}
			   \put(2,0){\line(0,1){2}}
			   \end{picture}}}
		      \,}

\newcommand*{\Next}{\!\raisebox{-.2ex}{ 
			\mbox{\unitlength=0.9ex%
			\begin{picture}(2,2)%
			\linethickness{0.06ex}%
			\put(1,1){\circle{2}} 
			\end{picture}}}       
			\,}

\def\start{\hbox{\bf start}}

\let\imp=\Rightarrow

\renewcommand{\TRPppV}{TeMP\xspace}
\title{Efficient First-Order Temporal Logic for Infinite-State Systems}
\author{Clare Dixon\quad Michael Fisher\quad Boris Konev\quad Alexei 
  Lisitsa\\[1ex]
Department of Computer Science, University of Liverpool\\
Liverpool L69 3BX, United Kingdom\\[1ex]
{\normalsize\texttt{\{C.Dixon,$\;$M.Fisher,$\;$B.Konev,$\;$A.Lisitsa\}@csc.liv.ac.uk}}}
\date{  }
\begin{document}
\maketitle
\begin{abstract}
In this paper we consider the specification and verification of
infinite-state systems using temporal logic. In particular, we
describe parameterised systems using a new variety of first-order
temporal logic that is both powerful enough for this form of
specification and tractable enough for practical deductive
verification. Importantly, the power of the temporal language allows
us to describe (and verify) asynchronous systems, communication delays
and more complex properties such as liveness and fairness
properties. These aspects appear difficult for many other approaches
to infinite-state verification.
\end{abstract}

\section{Introduction}
First-order temporal logic (\fotl) has been shown to be a powerful
formalism for expressing sophisticated dynamic
properties. Unfortunately, this power also leads to strong
intractability. Recently, however, a fragment of \fotl{}, called
\emph{monodic} \fotl{}, has been investigated, both in terms of its
theoretical~\cite{HWZ00,Hodkinson:Packed} and
practical~\cite{DFK03:ToCL,KDDFH05:IC,HKRV04:IJCAR} properties.
Essentially, monodicity allows for one free variable in every temporal
formula. Although clearly restrictive, this fragment has been shown to
be useful in expressive description logics, infinite-state
verification, and spatio-temporal
logics~\cite{AFWZ02,SturmW02,Kontchakovetal02,GKKWZ03,FKL06:VISSAS}.

We here develop a new temporal logic, combining decidable fragments of
monodic \fotl{}~\cite{HWZ00} with recent developments in XOR temporal
logics~\cite{DFK07:IJCAI}, and apply this to the verification of
parameterised systems. We use a communicating finite state machine
model of computation, and can specify not only basic synchronous,
parameterised systems with instantaneous broadcast
communication~\cite{esparza99verification}, but the powerful temporal
language allows us also to specify asynchronously executing machines
and more sophisticated communication properties, such as delayed
delivery of messages. In addition, and in contrast to many other
approaches~\cite{maidl01unifying,Del03,AbdullaJRS06}, not only safety,
but also liveness and fairness properties, can be verified through
automatic deductive verification. Finally, in contrast to work on
regular model checking~\cite{AbdullaJNdS04} and constraint based
verification using counting abstraction~\cite{esparza99verification},
the logical approach is both complete and decidable.

The verification of concurrent systems often comes down to the
analysis of multiple finite-state automata, for example of the following
form. 
\begin{center}
\includegraphics[width=0.25\textwidth]{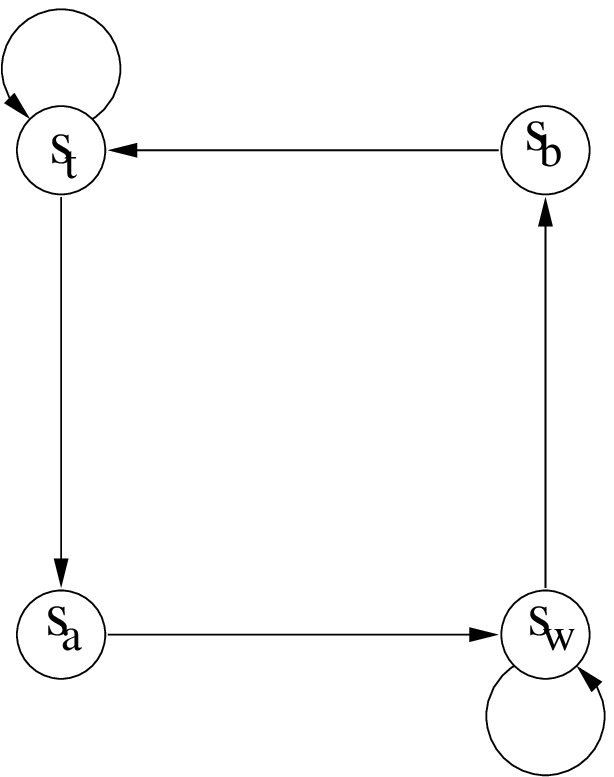}
\end{center}
In describing such automata, both automata-theoretic and logical
approaches may be used. While \emph{temporal logic}~\cite{emerson:90a}
provides a clear, concise and intuitive description of the system,
automate-theoretic techniques such as \emph{model
  checking}~\cite{Clarke00:MC} have been shown to be more useful in
practice. Recently, however, a propositional, linear-time temporal
logic with improved deductive properties has been
introduced~\cite{DFK06:TIME,DFK07:IJCAI}, providing the possibility of
practical deductive verification in the future. The essence of this
approach is to provide an XOR constraint between key
propositions. These constraints state that exactly one proposition from a XOR
set can be true at any moment in time. Thus, the automaton above can be
described by the
following clauses which are implicitly in the scope of a `$\always$' (`always
in the future') operator. 
$$
\begin{array}{ll}
1.  & \start\imp s_t\\
2.  & s_t\imp \Next(s_t\lor s_a)\\
3.  & s_b\imp \Next s_t\\
4.  & s_a\imp \Next s_w\\
5.  & s_w\imp \Next(s_w\lor s_b)
\end{array}
$$
Here `$\next$' is a temporal operator denoting `at the next moment' and 
`$\start$' is a temporal operator which holds only at the initial moment in
time. The inherent assumption that at any moment in time exactly one of 
$s_a$, $s_b$, $s_t$ or $s_w$ holds, is denoted by the following.
$$
\always(s_a \oplus s_b \oplus s_t \oplus s_w)
$$
With the complexity of the decision problem (regarding $s_a$, $s_b$,
etc) being polynomial, then the properties of any finite collection of
such automata can be tractably verified using this
\emph{propositional} XOR temporal logic.

However, one might argue that this deductive approach, although
elegant and concise, is still no better than a model checking
approach, since it targets just \emph{finite} collections of (finite)
state machines. Thus, this naturally leads to the question of whether
the XOR temporal approach can be extended to \emph{first-order
  temporal logics} and, if so, whether a form of tractability still
applies. In such an approach, we can consider \emph{infinite} numbers
of finite-state automata (initially, all of the same
structure). Previously, we have shown that \fotl{} can be used to
elegantly specify such a system, simply by assuming the argument to
each predicate represents a particular
automaton~\cite{FKL06:VISSAS}. Thus, in the following $s_a(X)$ is true
if automaton $X$ is in state $s_a$:
$$
\begin{array}{ll}
1.  & \start\imp \exists x. s_t(x)\\
2.  & \forall x.\ (s_t(x)\imp \Next(s_t(x)\lor s_a(x)))\\
3.  & \forall x.\ (s_b(x)\imp \Next s_t(x))\\
4.  & \forall x.\ (s_a(x)\imp \Next s_w(x))\\
5.  & \forall x.\ (s_w(x)\imp \Next(s_w(x)\lor s_b(x)))
\end{array}
$$
Thus, \fotl{} can be used to specify and verify broadcast protocols
between synchronous components~\cite{esparza99verification}. In this
paper we define a logic, \fotlx, which allows us to not only to
specify and verify systems of the above form, but also to specify and
verify more sophisticated asynchronous systems, and to carry out
verification with a reasonable complexity.

\section{\bffotlx}

\subsection{First-Order Temporal Logic}
First-Order (discrete, linear time) Temporal Logic, \fotl, is an
extension of classical first-order logic with operators that deal with
a discrete and linear model of time (isomorphic to the Natural
Numbers, $\Nat$).

\paragraph{Syntax.}
The symbols used in \fotl{} are
\begin{itemize}
\item \emph{Predicate symbols:} $P_0, P_1,\dots$ each of which is of a
  fixed arity (null-ary predicate symbols are \emph{propositions});
\item \emph{Variables:} $x_0, x_1,\dots$;
\item \emph{Constants:} $c_0,c_1,\dots$;
\item \emph{Boolean operators:} $\land$, $\lnot$, $\lor$, $\implies$,
  $\equiv$, $\true$ (`true'), $\false$ (`false');
\item \emph{First-order Quantifiers:} $\forall$ (`for all') and
  $\exists$ (`there exists'); and
\item \emph{Temporal operators:} $\always$ (`always in the future'),
  $\sometime$ (`sometime in the future'), $\next$ (`at the next
  moment'), $\until$ (until),  $\unless$ (weak until), and $\start$ (at the
  first moment in time).
\end{itemize}
Although the language contains constants, neither equality nor
function symbols are allowed. 

The set of well-formed \fotl{}-formulae is defined in the standard
way~\cite{HWZ00,DFK03:ToCL}:
\begin{itemize}
\item Booleans $\true$ and $\false$ are atomic \fotl-formulae;
\item if $P$ is an $n$-ary predicate symbol and
$t_i$, $1\leq i \leq n$, are variables or 
constants, then $P(t_1,\dots,t_n)$ is an atomic \fotl-formula;
\item if $\phi$ and $\psi$ are \fotl-formulae, so are 
$\lnot \phi$, $\phi\land\psi$, $\phi\lor\psi$, $\phi\implies\psi$,
and $\phi\equiv\psi$; 
\item if $\phi$ is an \fotl-formula and $x$ is a variable, then 
$\forall x \phi$ and $\exists x \phi$ are \fotl-formulae;
\item if $\phi$ and $\psi$ are \fotl-formulae, then so are 
$\always\phi$, $\sometime\phi$, $\next\phi$, $\phi\until\psi$, 
$\phi\unless\psi$, and $\start$.
\end{itemize}
A \emph{literal} is an atomic \fotl-formula or its negation.

\paragraph{Semantics,}
Intuitively, \fotl formulae are interpreted in \emph{first-order
  temporal structures} which are sequences $\gM$ of \emph{worlds},
$\gM = \gM_0,\gM_1,\dots$ with truth values in different worlds being
connected via temporal operators.

More formally, for every moment of time $n\geq 0$, there is a corresponding 
\emph{first-order} structure, $\gM_n = \langle D_n, I_n\rangle$, where
every $D_n$ is a non-empty set such that whenever $n<m$, $D_n\subseteq  D_m$,
and $I_n$ is  an interpretation of predicate and constant symbols over $D_n$.
We require that the interpretation of constants is \emph{rigid}. Thus, for
every constant $c$ and all moments of time $i,j\geq 0$, we have $I_i(c)=
I_j(c)$.

A \emph{(variable) assignment} $\ga$ is a function from the set of individual
variables to $\cup_{n\in\Nat} D_n$. We denote the set of all assignments by 
$\gA$. The set of variable assignments $\gA_n$ corresponding to
$\gM_n$ is a subset of the set of all assignments,
$\gA_n = \{\ga\in\gA\;|\; \ga(x)\in D_n \textrm{ for every variable $x$}\}$; 
clearly, $\gA_n\subseteq\gA_m$ if $n<m$.

The \emph{truth} relation $\gM_n\models^\ga \phi$ in a structure $\gM$,
is defined inductively on the construction of $\phi$ 
\emph{only for those assignments $\ga$ that satisfy the condition
$\ga\in\gA_n$}. See Fig.~\ref{fig:sem} for details.
\begin{figure*}[t]
\begin{small}$$
\begin{array}{|lcl|}\hline
& &  \\ 
\gM_n\models^\ga \true&& \gM_n\not\models^\ga\false\\
\gM_n\models^\ga
\start &\textrm{iff} & n=0 \\
\gM_n\models^\ga P(t_1,\dots,t_m) & \textrm{iff} & 
\langle I_n^\ga(t_1),\dots I_n^\ga(t_m)\rangle \in I_n(P), \textrm{ where }\\
& &I_n^\ga(t_i)=I_n(t_i),
\textrm{ if $t_i$ is a constant, and } I_n^\ga(t_i) = \ga(t_i),
\textrm{ if $t_i$ is a variable}\\ 
\gM_n\models^\ga \lnot \phi & \textrm{iff}& \gM_n\not\models^\ga\phi\\
\gM_n\models^\ga \phi\land \psi & \textrm{iff}& \gM_n\models^\ga\phi\textrm{ and } \gM_n\models^\ga\psi\\
\gM_n\models^\ga \phi\lor \psi & \textrm{iff}& \gM_n\models^\ga\phi\textrm{ or } \gM_n\models^\ga\psi\\
\gM_n\models^\ga \phi\implies \psi & \textrm{iff}& \gM_n\models^\ga(\lnot\phi\lor\psi) \\
\gM_n\models^\ga \phi\equiv \psi & \textrm{iff}& \gM_n\models^\ga((\phi\implies\psi)\land(\psi\implies\phi)) \\
\gM_n\models^\ga \forall x \phi & \textrm{iff} & \gM_n\models^\gb\phi
\textrm{ for every assignment $\gb$ that may differ}~\textrm{from
  $\ga$ only in $x$ and such that $\gb(x)\in D_n$}\\ 
\gM_n\models^\ga \exists x \phi & \textrm{iff} & \gM_n\models^\gb\phi
\textrm{ for some assignment $\gb$ that may differ from $\ga$ only in
  $x$ and such that $\gb(x)\in D_n$}\\ 
\gM_n\models^\ga\next\phi & \textrm{iff}& \gM_{n+1}\models^\ga\phi;\\
\gM_n\models^\ga\sometime\phi & \textrm{iff}& \textrm{there exists } m\geq n \textrm{ such that } \gM_{m}\models^\ga\phi; \\
\gM_n\models^\ga\always\phi & \textrm{iff}& \textrm{for all $m\geq n$, } \gM_m\models^\ga\phi;\\
\gM_n\models^\ga(\phi\until\psi) & \textrm{iff}& \textrm{there exists $m\geq
n$, such that } \gM_m\models^\ga\psi\textrm{ and, for all } i\in\Nat,
n\leq i < m \textrm{ implies } \gM_i\models^\ga\phi;\\ 
\gM_n\models^\ga(\phi\unless\psi) & \textrm{iff }&
\gM_n\models^\ga(\phi\until\psi)\textrm{ or }
\gM_n\models^\ga\always\phi.\\ 
\mbox{\ } & & \\
\hline
\end{array}
$$
\end{small}
\caption{Semantics of \fotl.}\label{fig:sem}
\end{figure*}
$\gM$ is a \emph{model} for a formula $\phi$ (or $\phi$ is \emph{true} in
$\gM$) if, and only if, there exists an assignment $\ga$ in $D_0$ such that
$\gM_0\models^\ga\phi$.  A formula is \emph{satisfiable} if, and only if,
it has a model. A formula is \emph{valid} if, and only if, it is true in any
temporal structure $\gM$ under any assignment $\ga$ in $D_0$.

\New%
The models introduced above are known as \emph{models with expanding
  domains} since $D_n\subseteq D_{n+1}$.  Another important class of
models consists of \emph{models with constant domains} in which the
class of first-order temporal structures, where \fotl{} formulae are
interpreted, is restricted to structures $\gM = \langle D_n,
I_n\rangle$, $n\in\Nat$, such that $D_i = D_j$ for all
$i,j\in\Nat$. The notions of truth and validity are defined similarly
to the expanding domain case.  It is known~\cite{WZ01DecModal} that
satisfiability over expanding domains can be reduced to satisfiability
over constant domains with only a polynomial increase in the size of
formulae.

\subsection{Monodicity and Monadicity}
The set of valid formulae of \fotl{} is not recursively
enumerable. Furthermore, it is known that even ``small'' fragments of
\fotl, such as the \emph{two-variable monadic} fragment (where all
predicates are unary), are not recursively
enumerable~\cite{Merz:Incomp:1992,HWZ00}.  However, the set of valid
\emph{monodic} formulae is known to be finitely
axiomatisable~\cite{WZ:APAL:AxMono}.
\begin{definition}
An \fotl-formula $\phi$ is 
called \emph{monodic} if, and only if, any subformula of the form $\mT\psi$,
where $\mT$ is one of $\next$, $\always$, $\sometimes$ (or $\psi_1\mT\psi_2$,
where $\mT$ is one of $\until$, $\unless$), contains at most one free variable.
\end{definition}
We note that the addition of either equality or function symbols to
the monodic fragment generally leads to the loss of recursive
enumerability~\cite{WZ:APAL:AxMono,DFL02:StudiaLogica,Hodkinson:Packed}.
Thus, monodic \fotl{} is expressive, yet even small extensions lead to
serious problems. Further, even with its recursive enumerability,
monodic \fotl{} is generally undecidable. To recover decidability, the
easiest route is to restrict the first order part to some decidable
fragment of first-order logic, such as the guarded, two-variable or
monadic fragments. We here choose the latter, since monadic predicates
fit well with our intended application to parameterised
systems. Recall that monadicity requires that all predicates have
arity of at most `1'. Thus, we use monadic, monodic
\fotl{}~\cite{DFK03:ToCL}.

A practical approach to proving monodic temporal formulae is to use
\emph{fine-grained temporal resolution}~\cite{KDDFH05:IC}, which has
been implemented in the theorem prover \TRPppV~\cite{HKRV04:IJCAR}.
In the past, \TRPppV{} has been successfully applied  to problems from
several domains~\cite{GHDFK05:JAR}, in particular, to examples specified in
the temporal 
logics of knowledge (the fusion of propositional linear-time temporal
logic with multi-modal S5)~\cite{DFW97:JLC,Dix05,DFK06:TIME}.
 From this work it is clear that monodic first-order temporal logic is
an important tool for specifying complex systems. However, it is also
clear that the complexity, even of \emph{monadic} monodic first-order
temporal logic, makes this approach difficult to use for larger
applications~\cite{GHDFK05:JAR,FKL06:VISSAS}.      

\subsection{XOR Restrictions}
An additional restriction  we make to the above logic involves implicit XOR
constraints 
over predicates. Such restrictions were introduced into temporal
logics in~\cite{DFK06:TIME}, where the correspondence with B{\"u}chi
automata was described, and generalised in~\cite{DFK07:IJCAI}. In both
cases, the decision problem is of much better (generally, polynomial)
complexity than that for the standard, unconstrained, logic.
However, in these papers only \emph{propositional} temporal logic was
considered. We now add such an XOR constraint to \fotlx{}.

The set of predicate symbols $\Pi=\{P_0, P_1,\dots\}$, is now partitioned
into a set of XOR-sets, $X_1$, $X_2$, $\ldots$, $X_n$, with one
\emph{non-XOR} set $N$ such that
\begin{enumerate}
\item all $X_i$ are disjoint with each other,
\item $N$ is disjoint with every $X_i$,
\item $\Pi$\ $=\ \displaystyle\bigcup_{j=0}^n X_j \,\cup\, N$, and
\item for each $X_i$, exactly \emph{one} predicate within $X_i$ is
  satisfied (for any element of the domain) at any moment in time.
\end{enumerate}
\begin{example}
Consider the formula
$$
\forall x.\ ((P_1(x)\lor P_2(x))\land (P_4(x)\lor P_7(x)\lor P_8(x)))
$$
where $\{P_1,P_2\}\subseteq X_1$ and  $\{P_4,P_7,P_8\}\subseteq
X_2$. The above formula states that, for any element of the domain, a,
then one of $P_1(a)$ or $P_2(a)$ must be satisfied and one of
$P_4(a)$, $P_7(a)$ or $P_8(a)$ must be satisfied.
\end{example}

\subsection{Normal Form}
%
To simplify our description, we will define a \emph{normal form} into
which \fotlx{} formulae can be translated.  In the following:
\begin{itemize}
\item $\accentset{\land}{X}^{\minus}_{ij}(x)$ denotes a conjunction of
  negated XOR predicates from the set $X_i$;
\item $\accentset{\lor}{X}^{\plus}_{ij}(x)$ denotes a disjunction of
  (positive) XOR predicates from the set $X_i$;
\item $\accentset{\land}{N}_i(x)$ denotes a conjunction of non-XOR
  literals;
\item $\accentset{\lor}{N}_i(x)$ denotes a disjunction of non-XOR
  literals.
\end{itemize}
A \emph{step} clause is defined as follows:
$$
\begin{array}{l}
\accentset{\land}X^\minus_{1j}(x) \land \ldots \accentset{\land}X^\minus_{nj}(x)
\land \accentset{\land}N_j(x)
\imp \\
\hspace*{2em}\next (\accentset{\lor}X^\plus_{1j}(x)  \lor \ldots \lor
\accentset{\lor}X^\plus_{nj}(x) \lor \accentset{\lor}N_j(x) )
\end{array}
$$
A \emph{monodic temporal problem in Divided Separated Normal Form
  (DSNF)}~\cite{DFK03:ToCL} is a quadruple $\langle \uP, \iP, \sP,
\eP\rangle$, where:
\begin{enumerate}
\itemsep=0pt
\item the universal part, $\uP$, is a finite set of arbitrary closed
  first-order formulae;
\item the initial part, $\iP$, is, again, a finite set of arbitrary
  closed first-order formulae;
\item the step part, $\sP$, is a finite set of step clauses; and
\item the eventuality part, $\eP$, is a finite set of 
eventuality clauses of the form $\sometime L(x)$, where $L(x)$ is a unary
literal.  
\end{enumerate}
In what follows, we will not distinguish between a finite set of
formulae ${\mathcal X }$ and the conjunction $\bigwedge {\mathcal X}$
of formulae within the set. With each  monodic temporal
problem, we associate the formula
$$
\iP\land\always\uP\land\always\forall x\sP\land\always\forall x\eP.
$$
Now, when we talk about particular properties of a temporal problem (e.g.,
satisfiability, validity, logical consequences etc) we mean properties of
the associated formula.

Every monodic \fotlx formula can be translated to the normal form in 
satisfiability preserving way using a renaming and unwinding
technique which substitutes non-atomic subformulae and replaces temporal
operators by their fixed point definitions as described, for example,
in~\cite{fdp01}. A step in this transformation  is the following: We
recursively rename each innermost open subformula $\xi(x)$, whose main
connective is a temporal operator, by $P_{\xi}(x)$, where $P_{\xi(x)}$ is a new
unary predicate, and rename each innermost closed subformula $\zeta$, whose
main connective is a temporal operator, by $p_{\zeta}$, where $p_{\zeta}$ is a
new propositional variable. While renaming introduces new, non-XOR predicates
and propositions, practical problems stemming from verification are nearly
in the normal form, see Section~\ref{sec:model}.

%
\subsection{Complexity}
First-order temporal logics are notorious for being of a high
complexity. Even decidable sub-fragments of monodic first-order
temporal logic can be too complex for practical use. For example,
satisfiability of monodic monadic \fotl{} logic is known to be
$\mathsf{EXPSPACE}$-complete~\cite{HKKWZ03}. However, imposing
XOR restrictions we obtain better complexity bounds.
\begin{theorem}\label{th:complexity}
Satisfiability of monodic monadic \fotlx{} formulae (in the normal
form) 
can be decided in $2^{O(N_1 \cdot N_2 \cdot\dots\cdot N_n\cdot 2^{N_a})}$ time,
where $N_1$,\dots, $N_n$ are cardinalities of the sets of XOR predicates, and
$N_a$ is the cardinality of the set of non-XOR predicates.
\end{theorem}
Before we sketch the proof of this result, we show how the XOR restrictions
influence the complexity of the satisfiability problem for monadic first-order
(non-temporal) logic.
\begin{lemma}\label{lemma:fom}
Satisfiability of monadic first-order formulae can be decided in
$\mathsf{NTime}(O(n\cdot{N_1 \cdot N_2 \cdot\dots\cdot N_n\cdot 2^{N_a}}))$,
where $n$ is the length of the formula, and $N_1$,\dots, $N_n$, $N_a$ are as in
Theorem~\ref{th:complexity}.
\end{lemma}
\begin{proof}
As in \cite{BGG97}, Proposition 6.2.9, the non-deterministic decision procedure
first guesses a structure and then verifies that the structure is a model for
the given formula. 
It was shown,
\cite{BGG97}, Proposition 6.2.1, Exercise 6.2.3, that if a monadic first-order
formula has a model, it also has a model, whose domain is the set of all
\emph{predicate colours}.
A {predicate colour}, $\gamma$, is a set of unary literals such that for
every predicate $P(x)$ from the set of all predicates $X_1\union\dots,
X_n\union N$, either $P(x)$ or $\lnot P(x)$ belongs to $\gamma$.   Notice that under the conditions of the lemma, there are at
most ${N_1 \cdot N_2 \cdot\dots\cdot N_n\cdot 2^{N_a}}$ different predicate
colours. Hence, the structure to guess is of $O({N_1 \cdot N_2 \cdot\dots\cdot
N_n\cdot 2^{N_a}})$ size.

It should be clear that one can evaluate a monadic formula of the size $n$ in a
structure of the size $O({N_1 \cdot N_2 \cdot\dots\cdot N_n\cdot
2^{N_a}})$ in deterministic $O(n\cdot {N_1 \cdot N_2 \cdot\dots\cdot N_n\cdot
2^{N_a}})$ time. Therefore, the overall complexity of the decision procedure is
$\mathsf{NTime}(O(n\cdot{N_1 \cdot N_2 \cdot\dots\cdot N_n\cdot 2^{N_a}}))$.
\end{proof}
\begin{proof}[of Theorem~\ref{th:complexity}, Sketch]
  For simplicity of presentation, we assume the formula contains no
  propositions.  Satisfiability of a monodic \fotl{} formula is
  equivalent to a property of the \emph{behaviour graph} for the
  formula, checkable in time polynomial in the product of the number of 
  different predicate colours and the size of the graph,
  see~\cite{DFK03:ToCL}, Theorem 5.15. For unrestricted \fotl{}
  formulae, the size of the behaviour graph is double exponential in
  the number of predicates. We estimate now the size of the behaviour
  graph and time needed for its construction for \fotlx{} formulae.

  Let $\Gamma$ be a set of predicate colours and $\rho$ be a map from the set
  of constants, $\const(\TProb)$, to $\Gamma$.  A couple $\langle\Gamma,
  \rho\rangle$ is called a \emph{colour scheme}.
  Nodes of the behaviour graph are {colour schemes}. Clearly, there
  are no more than $2^{O(N_1 \cdot N_2 \cdot\dots\cdot N_n\cdot
    2^{N_a})}$ different colour schemes. However, not every colour
  scheme is a node of the behaviour graph: a colour scheme
  $\mathcal{C}$ is a node if, and only if, a monadic formula of
  first-order (non-temporal) logic, constructed from the given
  \fotlx{} formula and the colour scheme itself, is satisfiable (for
  details see~\cite{DFK03:ToCL}). A similar first-order monadic condition
  determines which nodes are connected with edges. It can be seen that the size
  of the formula is polynomial in both cases.
 By Lemma~\ref{lemma:fom}, satisfiability of monadic first-order formulae can be
 decided in deterministic $2^{O(N_1 \cdot N_2 \cdot\dots\cdot N_n\cdot
 2^{N_a})}$ time.

  Overall, the behaviour graph, representing all possible models, for
  an \fotlx formula can be constructed in $2^{O(N_1 \cdot N_2
    \cdot\dots\cdot N_n\cdot 2^{N_a})}$ time.
\end{proof}


\section{Infinite-State Systems}
\label{sec:model}
%
%
In previous work,
notably~\cite{esparza99verification,delzanno00automatic} a
parameterised  finite state machine based model,
suitable for the specification and verification of protocols over
arbitrary numbers of processes was defined. Essentially, this uses a
family of identical, and synchronously executing, finite state
automata with a rudimentary form of communication: if one automaton
makes a transition (an action) $a$, then it is required that
\emph{all} other automata simultaneously make a complementary
transition (reaction) $\bar{a}$. In \cite{FKL06:VISSAS} we translated
this automata model into monodic \fotl{} and used automated theorem
proving in that logic to verify parameterised cache coherence
protocols~\cite{Del03}. The model assumed not only synchronous
behaviour of the communicating automata, but instantaneous broadcast.

Here we present a more general model suitable for specification of
both synchronous and asynchronous systems (protocols) with (possibly)
delayed broadcast and give its faithful translation into \fotlx. This
not only exhibits the power of the logic but, with the improved
complexity results of the previous section, provides a route towards
the practical verification of temporal properties of infinite state
systems.

\subsection{Process Model}
We begin with a description of both the asynchronous model, and the 
delayed broadcast approach.

\begin{definition}[Protocol]\label{def:protocol-simple}
A protocol, {\cal P} is a tuple $\langle Q, I, \Sigma,\tau \rangle$, 
where

\begin{itemize}
\item $Q$ is a finite set of states; 
\item $I \subseteq Q$ is a set of initial states;  
\item $\Sigma = \Sigma_{L} \cup \Sigma_{M} \cup \bar{\Sigma}_{M}$, where
\begin{itemize}
\item $\Sigma_{L}$ is a finite set of local actions; 
\item $\Sigma_{M}$ is a finite set of broadcast actions,\\ i.e. ``send a
  message''; 
\item $\bar{\Sigma}_{M} = \{\bar{\sigma}\mid \sigma \in \Sigma_{M}\}$ is 
the set of broadcast reactions, i.e. ``receive a message'';
\end{itemize}
\item $\tau \subseteq  Q \times \Sigma \times Q$ is a transition
relation that satisfies the following property 
\[\forall \sigma \in \Sigma_{M}.\ \forall q \in Q.\ \exists q' \in Q.\ 
\langle q,\bar{\sigma},q' \rangle \in \tau \] 
i.e., ``readiness to receive a message in any state''.
\end{itemize}
\end{definition}
Further, we define a notion of global machine, which is a set of $n$
finite automata, where $n$ is a parameter, each following the protocol
and able to communicate with others via (possibly delayed) broadcast.
To model asynchrony, we introduce a special automaton action,
$idle\,\not\in \Sigma$, meaning the automaton is not active and so its
state does not change. At any moment an arbitrary group of automata may
be idle and all non-idle automata perform their actions in accordance
with the transition function $\tau$; different automata may perform
different actions.

\begin{definition}[Asynchronous Global Machine]\label{def:glob_mach1} 
  Given a protocol, ${\cal P}= \langle Q, I, \Sigma,\tau \rangle$, the
  global machine ${\cal M}_{G}$ of dimension \emph{$n$} is the tuple $\langle
  Q_{{\cal M}_{G}}, I_{{\cal M}_{G}}
\tau_{{\cal M}_{G}}, {\cal E} \rangle$, where

\begin{itemize}
\item $Q_{{\cal M}_{G}} = Q^{n}$
\item $I_{{\cal M}_{G}} = I^{n}$
\item $\tau_{{\cal M}_{G}} \subseteq Q_{{\cal M}_{G}} 
\times (\Sigma \cup \{idle\}) ^{n} 
\times Q_{{\cal M}_{G}}$ is a transition
relation that satisfies the following property
$$\begin{array}{c}
\langle \langle s_{1}, \ldots, s_{n} \rangle, \langle \sigma_{1},
\ldots \sigma_{n} \rangle, \langle s'_{1}, \ldots, s'_{n} \rangle
\rangle \in \tau_{{\cal M}_{G}}\\
\hbox{\textit{iff}}\\
\forall 1\leq i\leq n.\ 
[(\sigma_{i} \not = idle \Rightarrow
\langle s_{i}, \sigma_{i}, s'_{i} \rangle \in \tau) \\
\land (\sigma_{i} = idle \Rightarrow s_{i} = s'_{i}]\,.
\end{array}$$

\item ${\cal E} = 2^{\Sigma_{M}}$ is a communication environment, that is a set of possible
sets of messages in transition.

\end{itemize} 
An element $G \in Q_{{\cal M}_{G}} \times (\Sigma \cup \{idle\}) ^{n}
\times {\cal E}$ is said to be a global configuration of the machine.

A run of a global machine ${\cal M}_{G}$ is a possibly infinite
sequence $\langle s^{1}, \sigma^{1}, E_{1} \rangle \ldots \langle
s^{i}, \sigma^{i}, E_{i} \rangle \ldots $ of global configurations
of ${\cal M}_{G}$ satisfying the properties (1)--(6) listed
below. In this formulation we assume $s^{i} = \langle s_{1}^{i},
\ldots, s_{n}^{i}\rangle$ and $ \sigma^{i} = \langle \sigma_{1}^{i},
\ldots, \sigma_{n}^{i} \rangle$.

\begin{enumerate}

\item $s^{1} \in I^{n}$\\
  (``initially all automata are in initial states'');  

\item $E_{1} = \emptyset$\\
 (``initially there are no messages in  transition''); 

\item  $\forall i.\ \langle s^{i}, \sigma^{i}  , s^{i+1} \rangle \in
  \tau_{{\cal M}_{G}}$ \\
(``an arbitrary part of the automata can fire'';

\item $\forall a \in \Sigma_{M}.\; \forall i. \; \forall j.\, (
  (\sigma^{i}_{j} = a) \Rightarrow \forall k. \; \exists l \ge i. \;
  (\sigma^{l}_{k} = \bar{a}))$\\
  (``delivery to all participants is guaranteed'');

\item $\forall a \in \Sigma_{M}. \; \forall i. \; \forall j. \; 
[(\sigma^{i}_{j} = \bar{a}) \Rightarrow (a \in E_{i}) \lor \exists k. \; \sigma^{i}_{k} = a )]$  
(``one can receive only messages  kept by the environment, or sent at the 
same moment of time '')  
 
\end{enumerate} 
In order to formulate further requirements we introduce the following
notation:

\[ Sent_{i} = \{a \in \Sigma_{M} \mid \exists j.\ \sigma^{i}_{j} = a \}\] 

\noindent $Delivered_{k} =$
$$
\left\{\begin{array}{c|l}
                 & \exists i \le k.\ (a \in Sent_{i}) \ \land\\
a \in \Sigma_{M} & (\forall l.\  (i < l < k) \rightarrow a \not\in
Sent_{l})\  \land\\
                 & (\forall j. \exists l.\ (i \le l \le k) \land
                 (\sigma^{l}_{j} = \bar{a}))
\end{array}\right\}$$
Then, the last requirement the run should satisfy is
\begin{enumerate} 
\item[6.] $\forall i.\ E_{i+1} = (E_{i} \cup Sent_{i}) - Delivered_{i}$ 
\end{enumerate} 
\end{definition} 

\paragraph{Example: Asynchronous Floodset Protocol.} 
We illustrate the use of the above model by presenting the
specification of an asynchronous FloodSet protocol in our model. This
is a variant of the \emph{FloodSet algorithm with alternative decision
  rule} (in terms of \cite{Lynch96:book}, p.105) designed for solution
of the Consensus problem. 

\begin{figure}[ht]
\begin{center}
\includegraphics[width=0.4\textwidth]{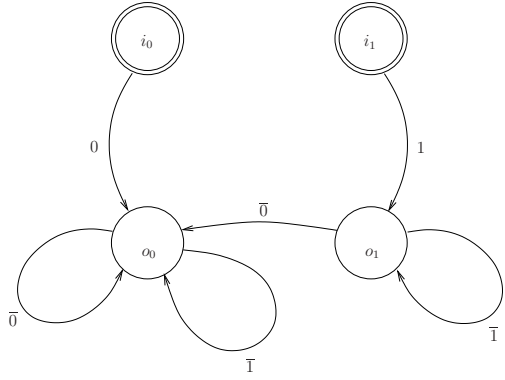}
\end{center}
\caption{Asynchronous FloodSet Protocol Process.\label{fig:flood}} 
\end{figure}

The setting is as follows. There are $n$ processes, each having an
{\em input bit} and an \emph{output bit}. The processes work
asynchronously, run the same algorithm and use {\em broadcast} for
communication. The broadcasted messages are guaranteed to be
delivered, though possibly with arbitrary delays. (The process is
described graphically in Fig.~\ref{fig:flood}.)

%

The goal of the algorithm is to eventually reach an agreement, i.e. to
produce an output bit, which would be the same for all 
processes. It is required also that if all
processes have the same input bit, that bit should be produced as an
output bit.

The asynchronous FloodSet protocol we consider here is adapted from 
\cite{Lynch96:book}. Main differences with original protocol are:

\begin{figure*}[t]
\begin{center}
\fbox{\begin{minipage}{0.98\textwidth}
\begin{itemize}

\item[{\bf I.}] Each automaton either performs one of the actions available
  in its state, or is idle:\\{}
 $\always[\forall x.\ P_{q}(x)
  \rightarrow A_{\sigma_{1}}(x)\lor \ldots\lor
  A_{\sigma_{k}}(x) \lor A_{idle}(x)]$, where $\{\sigma_{1}, \ldots
  \sigma_{k}\} = \{\sigma \in \Sigma \mid \exists r \langle q, \sigma,
  r \rangle \in \tau \}$.

\item[{\bf II.}] Action effects (non-deterministic actions):\quad$\always[\forall x P_{q}(x) \land A_{\sigma}(x)
  \rightarrow \next \bigvee_{\langle q, \sigma, r \rangle \in \tau}
  P_{r}(x)]$ for all $q \in S$ and $\sigma \in \Sigma$.

\item[{\bf III.}] Effect of being idle:\quad $\always[\forall x P_q(x) \land
  A_{idle}(x) \rightarrow \next P_q(x)]$, for all $q \in S$

\item[{\bf IV.}] Initially there are no messages in the transition and all
automata are in initial states:\quad
  $\start \rightarrow \neg m_{\sigma}$ for all $\sigma \in \Sigma_{m}$ and 
  $\start \rightarrow \forall x \bigvee_{q \in I} P_{q}(x)$. 

\item[{\bf V.}] All messages are eventually received (Guarantee of
  Delivery):\quad$\always [\exists y A_{\sigma}(y) \rightarrow
  \forall x \sometimes A_{\bar{\sigma}}(x)]$, for all $\sigma \in
  \Sigma_{m}$. 

\item[{\bf VI.}] Only messages kept in the environment (are in transition),
  or sent at the same moment of time can be received:\quad
  $\always[\forall x A_{\bar{\sigma}}(x) \rightarrow  m_{\sigma} \lor
  \exists y A_{\sigma}(y)   ]$ for all $\sigma \in \Sigma_{m}$. 

\item[{\bf VII.}]  Finally, for all $\sigma \in \Sigma_{m}$, we have
  the conjunction of the following formulae:

\begin{enumerate}
\item $\start \rightarrow \forall x.\ \neg Received_{\sigma}(x)$
\item $\always[\forall x.\  (A_{\bar{\sigma}}(x) \land \neg \forall
  y.\ 
Received_{\sigma}(y)) \rightarrow \next Received_{\sigma}(x)]$
\item $\always[\forall x.\  (Received_{\sigma}(x) \land \neg \forall
  y.\ Received_{\sigma}(y)\rightarrow \next Received_{\sigma}(x)]$
\item $\always[\forall x.\ (\neg (A_{\bar{\sigma}}(x) \lor
  Received_{\sigma}(x)) \land \neg \forall y.\ Received_{\sigma}(y))
  \rightarrow \next \neg Received_{\sigma}(x)]$
\item $\always[\forall x.\  Received_{\sigma} \rightarrow \next \neg m_{\sigma}]$
\item $\always[\exists x.\ A_{\sigma}(x) \land \neg \forall y.\ 
Received_{\sigma}(y) \rightarrow \next m_{\sigma}]$
\item $\always[\neg \exists x.\ A_{\sigma}(x) \land \neg \forall y.\ 
Received_{\sigma}(y) \rightarrow (m_{\sigma}\leftrightarrow 
\next m_{\sigma}] $
\end{enumerate}

\end{itemize}
\end{minipage}}
\end{center}
\caption{Temporal Specification of Abstract Protocol
  Structure.\label{fig:trans}} 
\end{figure*}

\begin{itemize}
\item the original protocol was synchronous, while our variant is asynchronous;
\item the original protocol assumed instantaneous message delivery,
  while we
  allow arbitrary delays in delivery; and 
\item although the original protocol was designed to work in the
  presence of crash (or fail-stop) failures, we assume, for
  simplicity, that there are no failures.
\end{itemize}
Because of the absence of failures the protocol is very simple and
unlike the original one does not require ``retransmission'' of any
value.  We will show later (in Section~\ref{sec:var}) how to include
the case of crash failures in the specification (and verification).
Thus, the asynchronous FloodSet protocol is defined, informally, as
follows.
\begin{itemize}
\item At the first round of computations, every process broadcasts its
      input bit.
\item At every round the (tentative) output bit is set to the
minimum value ever seen so far.
\end{itemize}
The correctness criterion for this protocol is that, eventually, the
output bits of all processes will be the same.
\medskip

\noindent 
Now we can specify the asynchronous FloodSet as a protocol $\langle
Q,I, \Sigma, \tau \rangle$, where $Q = \{ i_{0}, i_{1}, o_{0},
o_{1}\}$; $I = \{i_{0}, i_{1}\}$; $\Sigma = \Sigma_{m} \cup
\bar{\Sigma}_{m} \cup \Sigma_{L}$ with $\Sigma_{m} = \{0,1\}$,
$\bar{\Sigma}_{m} = \{\bar{0}, \bar{1}\}$, $\Sigma_{L} =
\emptyset$. The transition relation $\tau = \{\langle
i_{0},0,o_{0}\rangle,$ $\langle o_{0}, \bar{0}, o_{0}\rangle,$
$\langle o_{0}, \bar{1}, o_{0}\rangle,$ $\langle
i_{1},1,o_{1}\rangle,$ $\langle o_{1},\bar{0},o_{0}\rangle,$ $\langle
o_{1},\bar{1},o_{1} \rangle \}$.

\subsection{Temporal Translation} 
%
%
Given a protocol ${\cal P} = \langle Q,I,\Sigma,\tau\rangle$, we
define its translation to \fotlx{} as follows.  \medskip

\noindent For each $q \in Q$, introduce a monadic predicate symbol
$P_{q}$ and for each $\sigma \in \Sigma \cup \{idle\}$ introduce a
monadic predicate symbol $A_{\sigma}$. For each $\sigma \in
\Sigma_{M}$ we introduce also a propositional symbol $m_{\sigma}$.

Intuitively,
elements of the domain in the temporal representation will represent
exemplars of finite automata, and the formula $P_{q}(x)$ is intended to
represent ``automaton x is in state $q$''. The formula $A_{\sigma}(x)$ is going
to represent ``automaton $x$ performs action $\sigma$''.    
Proposition $m_{\sigma}$ will denote the fact 
``message $\sigma$ is in transition'' (i.e. it has been sent but not all
participants have received it.)  

Because of intended meaning we define two XOR-sets: 
$X_{1} = \{P_{q}\mid q \in Q \}$ and $X_{2} = \{A_{\sigma} \mid \sigma \in 
\Sigma \cup \{idle\}\}$. 
All other predicates belong to the set of non-XOR predicates.
 \medskip

 \noindent We define the temporal translation of ${\cal P}$, called
 $T_{\cal P}$, as a conjunction of the formulae in
 Fig.~\ref{fig:trans}. Note that, in order to define the temporal
 translation of requirement (6) above, (on the dynamics of 
environment updates) we introduce the
unary predicate symbol $Received_{\sigma}$ for every $\sigma \in
\Sigma_{m}$.
\medskip

\noindent We now consider the correctness of the temporal
translation. This translation of protocol $\cal P$ is faithful in the
following sense.

\begin{mfproposition}
\label{prop:trans}
Given a protocol, ${\cal P}$, and a global machine, ${\cal M}_{G}$, of
dimension $n$, then any temporal model $M_{1}, M_{2}, \ldots$ of
$T_{\cal P}$ with the finite domain $c_{1}, \ldots c_{n}$ of size $n$
represents some run $\langle s^{1}, \sigma^{1}, E_{1} \rangle \ldots \langle
s^{i}, \sigma^{i}, E_{i} \rangle \ldots $  of ${\cal M}_{G}$ as
follows:

$\langle \langle s_{1}, \ldots, s_{n} \rangle, 
          \langle \sigma_{1}, \ldots, \sigma_{n} \rangle, 
          E\rangle$ is $i$-th configuration of the run iff 
$M_{i} \models P_{q_{1}}(c_{1}) \land \ldots
P_{q_{n}}(c_{n})$,   $M_{i} \models A_{\sigma_{1}}(c_{1}) \land \ldots
A_{\sigma_{n}}(c_{n})$  and $E = 
\{\sigma \in \Sigma_{m} \mid M_{i} \models m_{\sigma}\}$

%
Dually, for any run of ${\cal M}_{G}$ there is a temporal model of
$T_{\cal P}$ with a domain of size $n$ representing this run.
\end{mfproposition} 

\begin{proof} 
By routine inspection of the definitions of runs, temporal models and
the translation.
\end{proof}

\subsection{Variations of the model}\label{sec:var}

The above model allows various modifications and corresponding version
of Proposition~\ref{prop:trans} still holds.

\paragraph{Determinism.} The basic model allows non-deterministic
actions. To specify the case of deterministic actions only, one should
replace the ``Action Effects'' axiom in Fig.~\ref{fig:trans} by the
following variant:
\[
\always[\forall x.\ P_{q}(x) \land A_{\sigma}(x) \rightarrow \next P_{r}(x)]
\]   for all $\langle q, \sigma, r  \rangle \in \tau$

\paragraph{Explicit bounds on delivery.} In the basic mode, no
explicit bounds on delivery time are given. To introduce bounds one
has to replace the ``Guarantee of Delivery'' axiom with the following
one:

\[
\always [\exists y.\, A_{\sigma}(y) \rightarrow \forall x.\, \next A_{\bar{\sigma}}(x) \lor \next A_{\bar{\sigma}}(x) \lor \ldots 
\lor \next^{n} A_{\bar{\sigma}}(x)]
\]  for all $\sigma \in \Sigma_{m}$ and some $n$ (representing the
maximal delay).

\paragraph{Finite bounds on delivery.} One may replace the ``Guarantee of
  Delivery'' axiom with the following one  

\[
\always [\exists y.\ A_{\sigma}(y) \rightarrow \sometimes \forall x.\ Received_{\bar{\sigma}}(x)]
\]  for all $\sigma \in \Sigma_{m}$.

\paragraph{Crashes.} One may replace the ``Guarantee of Delivery''
axiom by an axiom stating that only the messages sent by normal
(non-crashed) participants will be delivered to all participants.
(See~\cite{FKL06:VISSAS} for examples of such specifications in a
\fotl{} context.)


\paragraph{Guarded actions.} One can also extend the model with
guarded actions, where action can be performed depending on global
conditions in global configurations.
\bigskip

\noindent Returning to the FloodSet protocol, one may consider a
variation of the asynchronous protocol suitable for resolving the
Consensus problem in the presence of \emph{crash failures}. We can
modify the above setting as follows. Now, processes may fail and, from
that point onward, such processes send no further messages. Note,
however, that the messages sent by a process {\em in the moment of
  failure} may be delivered to {\em an arbitrary subset} of the
non-faulty processes.

The goal of the algorithm also has to be modified, so only
\emph{non-faulty} processes are required to eventually reach an
agreement. Thus, the FloodSet protocol considered above is modified by
adding the  following rule: 
\begin{itemize}
\item At every round (later than the first), a process broadcasts any
  value  {\em the first time it sees it}.
\end{itemize}
Now, in order to specify this protocol the variation of the model with
crashes should be used. The above rule can be easily encoded in the
model and we leave it as an exercise for the reader.

An interesting point here is that the protocol is actually correct
under the assumption that \emph{only finitely many processes may
  fail.}  This assumption is automatically satisfied in our automata
model, but not in its temporal translation. Instead, one may use the
above \emph{Finite bounds on delivery} axiom to prove the
correctness of this variation of the algorithm.

\subsection{Verification}
Now we have all the ingredients to perform the verification of
parameterised protocols.  Given a protocol $\cal P$, we can translate
it into a temporal formula $T_{\cal P}$. For the temporal
representation, $\chi$ of a required correctness condition, we then
check whether $T_{P} \rightarrow \chi$ is valid temporal formula. If
it is valid, then the protocol is correct for all possible values of
the parameter (sizes).

Correctness conditions can, of course, be described using any legal
\fotlx{} formula. For example, for the above FloodSet protocol(s) we
have a liveness condition to verify:

\[
\sometimes (\forall x.\ o_{0}(x) \lor \forall x.\ o_{1}(x)) 
\]

or, alternatively

\[
\sometimes \left[ \begin{array}{l}
    (\forall x.\ \mathit{Non}\hbox{\rm -}\mathit{faulty}(x) \rightarrow o_{0}(x))\ \lor\\
    (\forall x.\ \mathit{Non}\hbox{\rm -}\mathit{faulty}(x) \rightarrow o_{1}(x))
    \end{array}\right]
\] 
in the case of a protocol  working in presence of processor crashes.  

While space precludes describing many further conditions, we just note
that, in~\cite{FKL06:VISSAS}, we have demonstrated how this approach
can be used to verify safety properties, i.e with $\chi = \always
\phi$. Since we have the power of \fotlx{}, but with decidability
results, we can also automatically verify fairness formulae of the
form $\chi = \always\sometime\phi$.


\section{Concluding Remarks}
In the propositional case, the incorporation of XOR constraints within
temporal logics has been shown to be advantageous, not only because of
the reduced complexity of the decision procedure (essentially,
polynomial rather than exponential;~\cite{DFK07:IJCAI}), but also
because of the strong fit between the scenarios to be modelled (for
example, finite-state verification) and the XOR
logic~\cite{DFK06:TIME}). The XOR constraints essentially allow us to
select a set of names/propositions that must occur exclusively. In the
case of verification for finite state automata, we typically consider the
automaton states, or the input symbols, as being represented by such
sets. Modelling a scenario thus becomes a problem of engineering
suitable (combinations of) XOR sets.
 
In this paper, we have developed an XOR version of \fotl{}, providing:
its syntax and semantics; conditions for decidability; and detailed
complexity of the decision procedure.  As well as being an extension
and combination of the work reported in both~\cite{DFK03:ToCL}
and~\cite{DFK07:IJCAI}, this work forms the basis for tractable
temporal reasoning over infinite state problems. In order to motivate
this further, we considered a general model concerning the
verification of infinite numbers of identical processes. We provide an
extension of the work in~\cite{FKL06:VISSAS}
and~\cite{AbdullaJNdS04,AbdullaJRS06}, tackling liveness properties of
infinite-state systems, verification of asynchronous infinite-state
systems, and varieties of communication within infinite-state systems.
In particular, we are able to capture some of the more complex aspects
of \emph{asynchrony} and \emph{communication}, together with the
verification of more sophisticated \emph{liveness} and \emph{fairness}
properties.

The work in~\cite{FKL06:VISSAS} on basic temporal specification such
as the above have indeed shown that deductive verification can here be
attempted but is expensive --- the incorporation of XOR provides
significant improvements in complexity.

\subsection{Related Work}

The properties of first-order temporal logics have been studied, for
example, in~\cite{HWZ00,HKKWZ03}. Proof methods for the monodic
fragment of first order-temporal logics, based on resolution or
tableaux have been proposed in
~\cite{DFK03:ToCL,KDDFH05:IC,Kontchakovetal02}.

Model checking for parameterised and infinite state-systems is
considered in~\cite{AbdullaJNdS04}. Formulae are translated into to a
B\"uchi transducer with regular accepting states. Techniques from
regular model checking are then used to search for models. This
approach has been applied to several algorithms verifying safety
properties and some liveness properties.

Constraint based verification using counting
abstractions~\cite{delzanno00automatic,Del03,esparza99verification},
provides complete procedures for checking safety properties of
broadcast protocols.  However, such approaches
\begin{itemize}
\item have theoretically non-primitive recursive upper bounds for
  decision procedures (although they work well for small, interesting,
  examples) --- in  our case the upper bounds are definitely
  primitive-recursive;

\item are not suitable (or, have not been used) for asynchronous systems
  with delayed broadcast --- it is not clear how to adapt these
  methods for such systems; and 

\item typically lead to undecidable problems if applied to liveness
  properties. 
\end{itemize}

\subsection{Future Work}

Future work involves exploring further the framework described in this
paper in particular the development of an implementation to prove
properties of protocols in practice. Further, we would like to see if we
can extend the range of systems we can tackle beyond the monodic fragment.

We also note that some of the variations we might desire to include in
Section~\ref{sec:var} can lead to undecidable fragments. However, for
some of these variations, we have correct although (inevitably) incomplete
methods, see~\cite{FKL06:VISSAS}. We wish to explore these boundaries
further.



\begin{thebibliography}{10}\setlength{\itemsep}{-.8891ex}

\bibitem{AbdullaJNdS04}
P. A. Abdulla, B. Jonsson, M. Nilsson, J. d'Orso, and M. Saksena.
\newblock {Regular Model Checking for LTL(MSO)}.
\newblock In {\em Proc. 16th International Conference on Computer Aided
  Verification (CAV)}, volume 3114 of {\em LNCS},
  pages 348--360. Springer, 2004.

\bibitem{AbdullaJRS06}
P. A. Abdulla, B. Jonsson, A. Rezine, and M. Saksena.
\newblock {Proving Liveness by Backwards Reachability}.
\newblock In {\em Proc. 17th International Conference on Concurrency Theory
  (CONCUR)}, volume 4137 of {\em LNCS}, pages
  95--109. Springer, 2006.

\bibitem{AFWZ02}
A. Artale, E. Franconi, F. Wolter, and M. Zakharyaschev.
\newblock A Temporal Description Logic for Reasoning over Conceptual Schemas
  and Queries.
\newblock In {\em Proc. European Conference on Logics in Artificial
  Intelligence (JELIA)}, volume 2424 of {\em LNCS}, pages 98--110. Springer, 2002.

\bibitem{BGG97}
E. B{\"o}rger, E Gr{\"a}del, and Yu. Gurevich.
\newblock {\em The Classical Decision Problem}.
\newblock Springer, 1997.

\bibitem{BDFL02:LPAR} J. Brotherston, A. Degtyarev, M. Fisher, and
  A. Lisitsa.  \newblock {Implementing Invariant Search via Temporal
    Resolution}.  \newblock In {\em Proc. International Conference on
    Logic for Programming, Artificial Intelligence, and Reasoning
    (LPAR)}, volume 2514 of {\em LNCS},
  pages 86--101. Springer Verlag, 2002.  

\bibitem{Clarke00:MC}
E. Clarke, O. Grumberg, and D. Peled.
\newblock {\em Model Checking}.
\newblock MIT Press, Dec. 1999.

\bibitem{DFK03:ToCL}
A. Degtyarev, M. Fisher, and B. Konev.
\newblock {Monodic Temporal Resolution}.
\newblock {\em ACM Transactions on Computational Logic}, 7(1):108--150, January
  2006. (\texttt{ arXiv:cs.LO/0306041})

\bibitem{DFL02:StudiaLogica}
A. Degtyarev, M. Fisher, and A. Lisitsa.
\newblock {Equality and Monodic First-Order Temporal Logic}.
\newblock {\em Studia Logica}, 72(2):147--156, Nov. 2002.

\bibitem{delzanno00automatic}
G. Delzanno.
\newblock Automatic Verification of Parameterized Cache Coherence Protocols.
\newblock In {\em Proc. 12th International Conference on Computer Aided Verification
  (CAV)}, volume 1855 of {\em LNCS}, pages 53--68, 2000.

\bibitem{Del03}
G. Delzanno.
\newblock Constraint-based verification of parametrized cache coherence
  protocols.
\newblock {\em Formal Methods in System Design}, 23(3):257--301, 2003.

\bibitem{Dix05}
C. Dixon.
\newblock {Using Temporal Logics of Knowledge for Specification and
  Verification--a Case Study}.
\newblock {\em Journal of Applied Logic}, 4(1): 50-78, 2006.



\bibitem{DFFH04:TIME}
C. Dixon, M.C. Fern\'{a}ndez-Gago, M. Fisher,  and W. van der Hoek.
\newblock {Using  Temporal Logics of Knowledge in the Formal Verification of
Security Protocols.}
\newblock In {\em Proc. International Symposium on Temporal Representation and
  Reasoning (TIME)}, pages 148--151, 2004. IEEE CS Press,


\bibitem{DFK06:TIME}
C. Dixon, M. Fisher, and B. Konev.
\newblock {Is There a Future for Deductive Temporal Verification?}
\newblock In {\em Proc. International Symposium on Temporal Representation and
  Reasoning (TIME)}, pages 11--18, 2006. IEEE CS Press.

\bibitem{DFK07:IJCAI}
C. Dixon, M. Fisher, and B. Konev.
\newblock {Tractable Temporal Reasoning}.
\newblock In {\em Proc. International Joint Conference on Artificial
  Intelligence (IJCAI)}. AAAI Press, 2007.

\bibitem{DFW97:JLC}
C. Dixon, M. Fisher, and M. Wooldridge.
\newblock {R}esolution for {T}emporal {L}ogics of {K}nowledge.
\newblock {\em Journal of Logic and Computation}, 8(3):345--372, 1998.

\bibitem{emerson:90a}
E. A. Emerson.
\newblock {T}emporal and {M}odal {L}ogic.
\newblock In J. van Leeuwen, editor, {\em Handbook of Theoretical Computer
  Science}, pages 996--1072. Elsevier, 1990.

\bibitem{esparza99verification} J. Esparza, A. Finkel, and R. Mayr.
  \newblock {On the Verification of Broadcast Protocols}.  \newblock
  In {\em Proc.\ 14th {IEEE} Symposium on Logic in Computer Science
    (LICS)}, pages 352--359. IEEE CS Press, 1999.

\bibitem{fdp01}
M. Fisher, C. Dixon, and M. Peim.
\newblock {Clausal Temporal Resolution}.
\newblock {\em ACM Transactions on Computational Logic}, 2(1):12--56, Jan.
  2001. (\texttt{  arXiv:cs.LO/9907032})

\bibitem{FKL06:VISSAS}
M. Fisher, B. Konev, and A. Lisitsa.
\newblock Practical Infinite-state Verification with Temporal Reasoning.
\newblock In {\em Verification of Infinite State Systems and Security}. IOS Press, January 2006.

\bibitem{GKKWZ03}
D. Gabelaia, R. Kontchakov, A. Kurucz, F. Wolter, and M. Zakharyaschev.
\newblock On the Computational Complexity of Spatio-Temporal Logics.
\newblock In {\em Proc. 16th International Florida Artificial
  Intelligence Research Society Conference (FLAIRS)}, pages 460--464. AAAI
  Press, 2003.

\bibitem{GHDFK05:JAR}
M.-C. F. Gago, U. Hustadt, C. Dixon, M. Fisher, and B. Konev.
\newblock {First-Order Temporal Verification in Practice}.
\newblock {\em {Journal of Automated Reasoning}}, 34(3):295--321, 2005.

\bibitem{Hodkinson:Packed}
I. Hodkinson.
\newblock {Monodic Packed Fragment with Equality is Decidable}.
\newblock {\em Studia Logica}, 72(2):185--197, 2002.

\bibitem{HKKWZ03}
I. Hodkinson, R. Kontchakov, A. Kurucz, F. Wolter, and M. Zakharyaschev.
\newblock On the Computational Complexity of Decidable Fragments of First-Order
  Linear Temporal Logics.
\newblock In {\em Proc. International Symposium on Temporal
  Representation and Reasoning (TIME)}, pages 91--98. IEEE CS Press, 2003.

\bibitem{HWZ00}
I. Hodkinson, F. Wolter, and M. Zakharyashev.
\newblock {Decidable Fragments of First-Order Temporal Logics}.
\newblock {\em Annals of Pure and Applied Logic}, 2000.


\bibitem{HKRV04:IJCAR}
U. Hustadt, B. Konev, A. Riazanov, and A. Voronkov.
\newblock {TeMP: A Temporal Monodic Prover}.
\newblock In {\em Proc. 2nd International Joint Conference on Automated Reasoning (IJCAR)},
  volume 3097 of {\em LNAI}, pages 326--330. Springer, 2004.

\bibitem{KDDFH05:IC}
B. Konev, A. Degtyarev, C. Dixon, M. Fisher, and U. Hustadt.
\newblock Mechanising First-Order Temporal Resolution.
\newblock {\em Information and Computation}, 199(1-2):55--86, 2005.

\bibitem{Kontchakovetal02}
R. Kontchakov, C. Lutz, F. Wolter, and M. Zakharyaschev.
\newblock {Temoralising Tableaux}.
\newblock {\em Studia Logica}, 76(1):91--134, 2004.

\bibitem{Lynch96:book}
N. Lynch. 
\newblock {\em Distributed Algorithms}.
\newblock Morgan Kaufmann Publishers, San Mateo, CA, 1996.

\bibitem{maidl01unifying}
M. Maidl.
\newblock A Unifying Model Checking Approach for Safety Properties of
  Parameterized Systems.
\newblock {\em LNCS}, 2102:311--323, 2001.

\bibitem{Merz:Incomp:1992}
S. Merz.
\newblock Decidability and Incompleteness Results for First-Order Temporal
  Logic of Linear Time.
\newblock {\em Journal of Applied Non-Classical Logics}, 2:139--156, 1992.


\bibitem{SturmW02}
H. Sturm and F. Wolter.
\newblock {A Tableau Calculus for Temporal Description Logic: the Expanding
  Domain Case}.
\newblock {\em Journal of Logic and Computation}, 12(5):809--838, 2002.

\bibitem{WZ01DecModal}
F. Wolter and M. Zakharyaschev.
\newblock {Decidable Fragments of First-Order Modal Logics}.
\newblock {\em Journal of Symbolic Logic}, 66:1415--1438, 2001.

\bibitem{WZ:APAL:AxMono}
F. Wolter and M. Zakharyaschev.
\newblock Axiomatizing the Monodic Fragment of First-Order Temporal Logic.
\newblock {\em Annals of Pure and Applied Logic}, 118(1-2):133--145, 2002.

\end{thebibliography}

\end{document}